# Role of dislocations in scattering of charge carriers in mechanical polishing damaged layers of silicon wafers

Vladimir Yuryev [1]

*Natural Science Center of General Physics Institute of the Russian
Academy of Sciences, 38 Vavilov Street, Moscow 117942, Russia
e-mail: vyuryev@kapella.gpi.ru*

(Submitted March 28, 2000)

**Experimental observations.** Different dependences of the detector signal ($I_{sig}$) on the sample photoexcitation power ($W_{ex}$) observed for mechanically and mechano-chemically polished sides of the same single-crystalline silicon wafer in the optical-beam-induced IR-light scattering (dark field) mode of the scanning mid-IR-laser microscope (the probe $CO_2$-laser wavelength $\lambda_p = 10.6$ μm, the exciting HeNe-laser wavelength $\lambda_{ex} = 633$ nm) [1] were recently reported in Ref. [1] (see Fig. 1). Usual for linear recombination second-power law [1,2] was found for the mechano-chemically polished side of the wafer in the optical-beam-induced scattering mode. In the optical-beam-induced IR-light absorption mode [1], a linear law was obtained for both wafer sides which is also in agreement with the theoretical prediction. Only for the mechanically polished side an "anomalous" cubic dependence was registered in the induced light-scattering mode which has not been explained thus far.[2] Now the explanation has been found.

**A role of dislocations.** Let us suppose that the impact of charged extended linear defects (dislocation lines) on the carrier scattering process prevails in the mechanical polishing damaged layer and the defect density $N_m$ (per unit area) is so high that the probe IR light cyclic frequency $\omega \ll 1/\tau$ ($\tau$ is the momentum relaxation time).

To understand the physical picture, let us consider the following simple model. Let us assume that the dimensions of the defects $L_x$ and $L_z$ satisfy the condition: $L_z \gg L_D \gg L_x$ ($L_D$ is the Debye length). In this case, as the probability of carrier scattering on such defects $P_{sc} \propto L_D L_z$ (the carrier velocity component along the defect ($z$) axis is implied to be negligible in comparison with its component normal to the $z$-axis), $\tau \propto (1/L_D L_z) \propto n^{1/2}$; $n$ is the free carrier concentration.

Keeping in mind the assumption made about $\omega$ and $\tau$ it may be written for the IR-light scattering intensity: $I_{sc} \propto |\delta\varepsilon|^2 \propto [\text{Im}(\varepsilon)]^2$ [1,2] where $\delta\varepsilon$ is a local variation of the dielectric function $\varepsilon$. In turn, $\text{Im}(\varepsilon) \propto (\omega_p)^2 \tau \propto n^{3/2}$ ($\omega_p$ is the plasma frequency) that agrees with the measured cubic dependences (hereinafter, the linear recombination is implied, i.e. $n \propto W_{ex}$).

More accurate calculations give the same law for the dependence of the light scattering intensity on the photoexcitation power for the case of free carrier scattering on charged linear monopoles. One can easily derive following Ref. [3]:

$$\mathrm{Im}(\varepsilon) \approx \frac{2^{9/2} n^{3/2} \varepsilon_L^{3/2} \varepsilon_0^{1/2} e\, kT}{\pi^{1/2} N_m q^2 m^{*1/2} \omega\, g(p)} \quad (1)$$

where

$$g(p) = (1+2p)\exp(p)\,\mathrm{erfc}(p^{1/2}) - 2p^{1/2}/\pi^{1/2}, \quad (2)$$

$$p = \frac{\hbar^2 e^2 n}{8 m^* \varepsilon_L \varepsilon_0 (kT)^2}. \quad (3)$$

---

[1] Supported by the Ministry of Science and Technologies of the Russian Federation grant No. 02.04.3.2.40. Э .24.
[2] Remark, that the same cubic law was previously obtained for mechanically polished Ge sample by the low-angle mid-IR-light scattering technique with the surface photoexcitation [1].

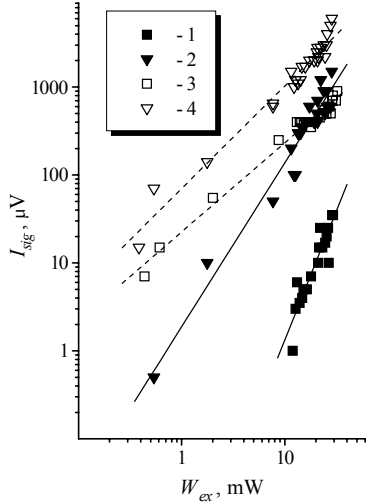

Figure 1. Mid-IR-laser microscope detector signal vs. the sample photoexcitation power in crystal for mechanically (1),(3) and mechano-chemically (2),(4) polished sides of the same CZ Si:B wafer; (1),(2): optical-beam-induced IR-light scattering (dark field); (3),(4): optical-beam-induced IR-light absorption; probe laser wavelength $\lambda_p = 10.6$ μm, exciting laser wavelength $\lambda_{ex} = 633$ nm; fitting line slope: (1) $2.9 \pm 0.3$, (2) $1.9 \pm 0.1$, (3) $1.01 \pm 0.03$, (4) $1.17 \pm 0.05$.

In eqs. (1) to (3), $\varepsilon_L$ is the lattice dielectric constant; $q$ is charge per defect unit length; $m^*$ is the carrier effective mass; the rest designations are used in their commonly known meaning.

It is seen from eqs. (1) to (3) that $\mathrm{Im}(\varepsilon) \propto n^{3/2}$ in a wide range of the free carrier concentrations—for Si or Ge crystals $g(p) \approx$ const up to $n \sim 10^{18}$ cm$^{-3}$ at $T = 300$ K.

So, it can be concluded from eq. (1) that the microscope detector signal

$$I_{sig} \propto I_{sc} \propto n^3 \propto W_{ex}^3. \quad (4)$$

This expression explains the cubic law observed for dependences of the light scattering intensity (detector signal) on the photoexcitation power obtained in the experiments for the mechanically polished silicon wafers (as well as the cubic law previously obtained for Ge).

**Conclusions.** The inference can be made from the above that dislocations play the main role in the mechanism of charge carrier scattering in silicon (or germanium) wafers subsurface layer damaged by mechanical polishing. The depth of this layer is estimated as comparable with or some greater than the space charge region width.